\documentstyle[psfig]{mn0}

\begin{document}

\def\fr2{$\frac{1}{2}$} \def\frc3{$\frac{1}{3}$} \def\f4{$\frac{1}{4}$} \def\fk8{$\frac{1}{8}$}
\def\x{{\rm x}} \def\n{{\rm \nu}} \def\e{{\rm e}} \def\el{{\rm el}} \def\B{{\rm B}} \def\c{{\rm c}}
\def\0{{\rm start}} \def\ended{{\rm end}} \def\sy{{\rm (syn)}} \def\ic{{\rm (iC)}} \def\d{{\rm d}}
\def\u51{$\times 10^{51}$} \def\y52{$\times 10^{52}$} \def\z53{$\times 10^{53}$} \def\w54{$\times 10^{54}$}
\def\p{{\rm p}} \def\s{{\rm s}} \def\b{{\rm b}} \def\i{{\rm i}} \def\o{{\rm o}} \def\r{\tilde{r}}
\def\j{{\rm j}} \def\1{{\rm 1}} \def\2{{\rm 2}} \def\ag{{\rm ag}} \def\rgamma{{\rm \gamma}}
\def\Egfag{E_{\rm \gamma}-f_{\rm ag}}
\def\lta{\mathrel{\rlap{\raise 0.511ex \hbox{$<$}}{\lower 0.511ex \hbox{$\sim$}}}}
\def\gta{\mathrel{\rlap{\raise 0.511ex \hbox{$>$}}{\lower 0.511ex \hbox{$\sim$}}}}

\parskip 3pt
\topmargin 0pt

\title[Energy injection in GRB afterglows]
  {The possible ubiquity of energy injection in Gamma-Ray Burst afterglows}

\author[Panaitescu \& Vestrand]{A. Panaitescu and W.T. Vestrand \\
       ISR-2: Space and Remote Sensing, MS B244, Los Alamos National Laboratory,
       Los Alamos, NM 87545, USA}

\maketitle

\begin{abstract}
\\
\begin{small}
 Since its launch in 2004, the Swift satellite has monitored the X-ray afterglows of several hundred
 Gamma-Ray Bursts, and revealed that their X-ray light-curves are more complex than previously thought,
 exhibiting up to three power-law segments. Energy injection into the relativistic blast-wave energizing
 the burst ambient medium has been proposed most often to be the reason for the X-ray afterglow complexity.
 We examine 117 light-curve breaks of 98 Swift X-ray afterglows, selected for their high-quality monitoring 
 and well-constrained flux decay rates. Thirty percent of afterglows have a break that can be an adiabatic 
 jet-break, in the sense that there is one variant of the forward-shock emission from a collimated outflow 
 model that can account for both the pre- and post-break flux power-law decay indices, given the measured 
 X-ray spectral slope. If allowance is made for a steady energy injection into the forward-shock, then 
 another 56 percent of X-ray afterglows have a light-curve break that can be explained with a jet-break. 
 The remaining 12 percent that are not jet-breaks, as well as the existence of two breaks in 19 afterglows
 (out of which only one can be a jet-break), suggest that some X-ray breaks arise from a sudden change 
 in the rate at which energy is added to the blast-wave, and it may well be that a larger fraction of 
 X-ray light-curve breaks are generated by that mechanism. 
 The fractional increase in the shock energy, inferred from the energy injection required to account 
 for the observed X-ray flux decays, may be anticorrelated with the GRB prompt output, whether
 the X-ray break is a jet-break or an energy-injection break. That anticorrelation can also be seen
 as bursts with a higher energy output being followed by faster fading X-ray afterglows.
 To test the above two mechanisms for afterglow light-curve breaks, we derive comprehensive analytical 
 results for the dynamics of outflows undergoing energy injection and for their light-curves, including 
 closure relations for inverse-Compton afterglows and for the emission from spreading jets interacting 
 with an wind-like ambient medium.
\end{small}
\\ \\ 
\end{abstract}


\section{Introduction}

 The power-law temporal decays displayed by the radio, optical, and X-ray fluxes of GRB afterglows
($F_\n \propto t^{-\alpha}$) have provided strong support for the prediction (Meszaros \& Rees 1997)
that afterglows arise from the relativistic blast-wave generated as the GRB outflow ejecta interacts
with the ambient medium. If the GRB outflow is a collimated jet then a break in the smooth power-law
decay of the afterglow is expected for both either a conical jet (Panaitescu, Meszaros \& Rees 1998)
or a sideways-spreading jet (by Rhoads 1999). The breaks found in the flux decay light curves of
most of the well-sampled afterglows that were localized by the BeppoSAX satellite seemed to provide
solid observational support for the existence of these predicted jet-breaks (e.g. Frail et al 2001).

 It was also realized early on that there should be a substantial diversity in the steepness
of the afterglow flux decay, even when the spectral slope is constant, if the fundamental 
forward-shock parameters vary as the shock traverses the ambient medium. Rees \& Meszaros (1998)
derived the expected power-law flux decay for the forward-shock synchrotron flux in the case of
an outflow whose kinetic energy-per-solid angle $dE/d\Omega$ is not constant in time, either 
because the shock's energy increases due to fresh ejecta arriving at the blast-wave or because
$dE/d\Omega$ is anisotropic. Two important microphysical parameters ($\epsilon_\e$ and $\epsilon_\B$),
which quantify the fraction of the post-shock energy acquired by electrons and magnetic field,
also determine the intensity the afterglow synchrotron flux, and their variability could also yield
a variety of decay indices $\alpha$.

 Swift measurements of GRB X-ray afterglows provide support for that shock parameters vary.
The standard forward-shock model 
with constant energy and shock micro-parameters predicts a closure relation between the temporal 
decay index $\alpha$ and the slope $\beta$ of the X-ray continuum ($F_\n \propto \nu^{-\beta}$) 
of the form $\alpha = 1.5 \beta + const$. The early X-ray flux of some Swift afterglows decays
slower than expected for the measured spectral slope $\beta$, which was attributed to a variable 
kinetic energy of the forward shock (Nousek et al 2006, Panaitescu et al 2006, Zhang et al 2006). 
Moreover, the flux decay indices $\alpha$ displayed by 
Swift X-ray afterglows after the slow decay phase are not correlated with the X-ray spectral 
slope $\beta$ (Panaitescu 2007), as would be expected from the $\alpha-\beta$ closure relation 
for the forward-shock emission.

 When optical light-curves are taken into account, the standard jet model fares even worse,
being inconsistent with practically all Swift afterglows (Liang et al 2008). While more than
half of the well-sampled optical afterglows display coupled optical and X-ray light-curves
(Panaitescu \& Vestrand 2011), exhibiting similar flux decay indices and/or simultaneous light-curve
breaks, many other afterglows show a decoupling of their optical and X-ray light-curves, with
a light-curve break occurring only in the X-ray.
At least four possibilities for this light-curve decoupling, which is not predicted by the standard
jet model, have been proposed:
$i$) the X-ray afterglow emission arises occasionally from the same not-yet-understood
    mechanism which produces the prompt burst emission (Ghisellini et al 2007),
$ii$) there is a dominant contribution of the reverse-shock to the afterglow emission
    (Uhm \& Beloborodov 2007),
$iii$) the forward-shock emission is reprocessed by bulk-scattering and/or inverse-Compton
    scattering outside the forward-shock (Panaitescu 2008), or
$iv$) the optical and X-ray afterglow flux arise from different parts of the forward-shock,
    from different jets (e.g. Racusin et al 2008).

 In this work, {\sl we assume that Swift X-ray afterglows arise from the forward-shock},
and attribute the failure of the standard jet model in accounting for the X-ray flux decay
indices to the assumption that the forward shock energy is constant. 

 Evolving micro-physical parameters (e.g. Ioka et al 2006) could, in principle, explain the
measured X-ray flux decay indices, but the existence of light-curve breaks requires an unnatural,
sudden change in their evolution. For that reason, we maintain in this work the basic assumption
of the standard jet model that micro-physical parameters do not evolve.

 We also assume that the ambient medium with which the post-GRB outflow interacts is uniform,
without discontinuities that could yield breaks in the afterglow light-curve at a frequency.
A notable model thus excluded is that where the afterglow-producing outflow interacts with a
prior outflow (Ioka et al 2006) whose density increases radially outward up to its termination
shock, yielding a slower decay of the afterglow flux than expected for a homogeneous ambient
medium, followed by a break to a steeper decay.

 One reason for relaxing the constant shock energy constraint is the observational evidence
for X-ray flares in Swift afterglows (e.g. Chincarini et al 2007). The short timescale of
those flares indicates that the GRB progenitor produces relativistic ejecta after the burst
phase, whose arrival at the decelerating forward-shock could increase substantially the
blast-wave's energy. Our hypothesis is that such an injection of energy in the forward-shock
is, in general, a more persistent and continuous process. However, we note that the fluence
of X-ray flares is about 10 percent that of the burst (Falcone et al 2007), hence the late
outflow powering the flares could carry a dynamically-important energy only if that outflow
is much less efficient than the GRB outflow in producing X-ray emission.

 We derive in \S\ref{anal} the closure relations $\alpha-\beta$ for a continuous energy
injection in the forward-shock. For the ease of extracting from observations the required
injected power history, we assume that the forward-shock energy increases as a power with
observer-time, $E \propto t^e$, which is motivated by that the resulting afterglow flux
should be power-law in observer time. This assumption circumvents a calculation of the
kinematics for the ejecta -- forward-shock interaction.

 In \S\ref{swift}, the closure relations for the forward-shock emission with energy injection
are applied to a set of Swift afterglows with well sampled light-curves. There, we examine 
two plausible scenarios for the origin of afterglow X-ray light-curve breaks (break in energy 
injection, jet-break), three dynamical regimes of the forward-shock (spherical expansion, 
conical jet, spreading jet), and two radiation processes that may yield the afterglow X-ray flux 
(synchrotron, inverse-Compton).

\section{Analytical flux decays for forward-shock models with energy injection}
\label{anal}

 There are three mechanisms that could produce a change in the flux power-law decay index $\alpha$,
i.e. an afterglow light-curve break. 

(1) Passage of a spectral break. X-ray light-curve breaks are, generally, not accompanied by 
a spectral evolution (Nousek et al 2006; Liang, Zhang \& Zhang 2007; Willingale et al 2007), 
hence their origin in 
the passage of the spectral feature is largely ruled out, but there are other reasons why 
such an X-ray light-curve break origin is unlikely. The slower evolution of the cooling frequency
$\nu_\c$ (the synchrotron characteristic frequency at which radiate the electrons that cool radiatively 
on the hydrodynamical timescale) leads to extended X-ray light-curve breaks when $\nu_\c$ crosses the
0.3-10 keV band. The peak frequency $\nu_\i$ of the $\nu F\n$ synchrotron spectrum decreases 
faster, but it is likely to be lower than the X-rays even at early times, hence it cannot cross
the X-ray band during afterglow observations.
 
(2) Jet-break. A light-curve break occurs at the `jet-break time', when the outflow Lorentz 
factor $\Gamma$ drops below the inverse of the jet half-angle $\theta_\j$. For a `conical jet' 
that does not spread laterally (as in the case of an outflow core embedded in an envelope that 
prevents the spreading of the core but yields a negligible emission), the steepening of the 
X-ray flux decay is solely due to that, after the jet-break time, the average kinetic energy
per solid angle $dE/d\Omega$ over the visible cone of angle $\theta = \Gamma^{-1}$ (where 
relativistic Doppler boost is effective) is decreasing with observer-frame time. 
For a `spreading jet' (as expected for a top-hat distribution of $dE/d\Omega$), in addition 
to the above-described geometrical-relativistic effect, there is a second contribution to the 
X-ray flux decay steepening from the enhanced jet deceleration produced by the jet lateral 
spreading. 

(3) Energy injection break.
If energy injection occurs routinely in GRB afterglows, then a light-curve break could
be produced by a change in the injected power. The dynamical behaviours of the forward-shock 
model to be considered are: 
$i$) a `spherical' outflow, i.e. a wide jet whose edge becomes visible to the observer after 
  the last X-ray measurement (at $t_\ended$), which occurs if jet half-angle satisfies 
  $\theta_\j > 1/\Gamma(t_\ended)$, and 
$ii$) a narrow jet whose edge is already visible at the first X-ray observation (at $t_\0$), 
  occurring when $\theta_\j < 1/\Gamma(t_\0)$. 
For the latter model, we will consider both a conical jet and a spreading jet.
These dynamical models are applicable also to X-ray afterglows without light-curve breaks, 
in which case the energy injection law must be unevolving (a single index $e$).

 We will investigate the ability of the energy-injection break and jet-break models above 
to accommodate the observed X-ray flux decays before and after the break, with allowance
for all three dynamical models (spherical outflow, conical jet, spreading jet), for two types 
of ambient medium (homogeneous and wind-like), two radiation mechanisms dominating the X-ray 
flux (synchrotron and inverse-Compton), and two possible locations of the X-ray relative to 
the cooling frequency $\nu_\c$ but, first, a word about their rationale.

 Wind-like media, having a $n \sim 0.1-1\, (r/10^{18}\,{\rm cm})^{-2}$ particle density, 
are expected at 0.1-1 pc from the 
source of long GRBs, i.e. where the $100\,{\rm s}-1\,{\rm Ms}$ afterglow emission is produced,
because the progenitors of such bursts are massive Wolf-Rayet stars that drive powerful winds. 
The termination-shock of such winds is of several pcs (Garcia-Segura et al 1996), thus a 
homogeneous medium at the afterglow location seems unjustified.
However, if the GRB progenitor resides in a highly-pressurized bubble, the wind termination-shock 
radius could be as low as 0.5 pc (Chevalier et al 2004), with a further reduction of that radius
if the WR wind is weak, as expected along the rotation axis (Ramirez-Ruiz et al 2005), or if 
the WR star moves sufficiently fast relative to the ISM (van Marle et al 2006), thus the 
afterglow could occur in shocked wind, which is quasi-homogeneous. 

 Synchrotron emission is generally believed to produce all the afterglow emission, but the
likelihood of inverse-Compton emission being dominant is higher at higher photon frequency,
like X-rays. Admittedly, that requires a high environment density, perhaps ten times higher
(Panaitescu 2011) than implied by the mass-loss rates and terminal wind velocities measured 
for Galactic Wolf-Rayet winds (Nugis \& Lamers 2000). 

\subsection{Spectral characteristics}

 The $\alpha-\beta$ closure relations expected for a given variant of the forward-shock model
are derived from the evolution of the spectral characteristics $F_\p$ (peak flux), $\nu_\i$ 
(frequency of the $\nu F_\n$ spectrum peak), and $\nu_\c$ (cooling frequency), and from the 
shape of the $F_\n$ spectrum at frequencies higher than its $\min(\nu_\i,\nu_\c)$ peak: 
$F_\n \propto \nu^{-1/2}$ 
for $\nu_\c < \nu < \nu_\i$, $F_\n \propto \nu^{-(p-1)/2}$ for $\nu_\i < \nu < \nu_\c$ and 
$F_\n \propto \nu^{-p/2}$ for $\max(\nu_\i,\nu_\c) < \nu$, where $p$ is the power-law index 
of the electron distribution with energy produced by the shock ($dN/d\varepsilon \propto 
\varepsilon^{-p}$). Therefore, the flux received at observing frequency $\nu$ satisfies 
$F_\n \propto F_\p \nu_\i^{(p-1)/2}$ for $\nu_\i < \nu < \nu_\c$ and $F_\n \propto F_\p 
\nu_\i^{(p-1)/2} \nu_\c^{1/2}$ for $\max(\nu_\i,\nu_\c) < \nu$ ($\nu_\c < \nu < \nu_\i$ implies
an X-ray spectrum with $\beta_\x = 1/2$ that is harder than measured for nearly all Swift X-ray 
afterglows).

 For a spherical outflow (or a jet before the jet-break time $t_\j$) and for synchrotron emission,
$F_\p \propto B M \Gamma$, $\nu_\i \propto \gamma_\i^2 B \Gamma$, $\nu_\c \propto \gamma_\c^2 B 
\Gamma$. Here $M$ is the mass of the ambient medium swept-up by the forward-shock (i.e. the 
number of radiating electrons); $B \propto (\epsilon_\B n)^{1/2} \Gamma$ is the post-shock 
magnetic field, whose energy density $B^2/(8\pi)$ is assumed to be a fraction $\epsilon_\B$ 
of the post-shock energy density $u' = 4\Gamma^2 n m_\p c^2$ (from the shock jump conditions); 
$n$ is the external density at the shock's location; $\gamma_\i \propto \epsilon_\e \Gamma$ 
is the typical electron random Lorentz factor, parametrized as a fraction $\epsilon_\e$
of the proton random Lorentz factor $\gamma_\p \simeq \Gamma$; and $\gamma_\c \propto 
[B^2 t' (Y+1)]^{-1}$ is the random Lorentz factor of the electrons whose radiative cooling
timescale equals the comoving-frame shock's age $t' = \int dr/(c\Gamma)$, $r$ being the 
outflow radius and $Y$ the Compton parameter. We assume the more likely situations of 
either electrons cooling mostly through synchrotron emission ($Y < 1$) or, if inverse-Compton
is dominant ($Y > 1$), that $\nu_\c < \nu_\i$, in which case $Y \simeq (\epsilon_\e/\epsilon_\B)^{1/2}$ 
is constant and does not affect the evolution of $\nu_\c$.

 For a jet (conical or spreading) after the jet-break time and for {\sl synchrotron} emission, 
we have $F_\p \propto B M \Gamma^3$, with $\nu_\i$ and $\nu_\c$ as above. Thus
\begin{equation}
  F_\p^\sy \propto \left\{ \hspace{-2mm} \begin{array}{ll} 
     \sqrt{n} M \Gamma^2 & (t< t_\j) \\ 
     \sqrt{n} M \Gamma^4 & (t> t_\j)  
     \end{array} \right.  \hspace{-2mm} 
\label{spsy1}
\end{equation}
\begin{equation}
  \nu_\i^\sy \propto \sqrt{n} \Gamma^4 
\end{equation}
\begin{equation}
  \nu_\c^\sy \propto (n^{3/2} \Gamma^2 t'^2)^{-1}
\label{spsy3}
\end{equation}

 The characteristics of the {\sl inverse-Compton} spectrum are immediately related with those of 
the synchrotron spectrum: $F_\p^\ic \simeq \tau_\el F_\p^\sy$ (where $\tau_\el \propto M/r^2$ is 
the optical thickness to electron scattering of the shocked ambient medium), $\nu_\i^\ic \simeq 
\gamma_\i^2 \nu_\i^\sy$, and $\nu_\c^\ic \simeq \gamma_\c^2 \nu_\c^\sy$, thus
\begin{equation}
  F_\p^\ic \propto \frac{M}{r^2} F_\p^\sy 
\label{spic1}
\end{equation}
\begin{equation}
  \nu_\i^\ic \propto \sqrt{n} \Gamma^6 
\end{equation}
\begin{equation}
  \nu_\c^\ic \propto (n^{7/2} \Gamma^6 t'^4)^{-1}
\label{spic3}
\end{equation}


\subsection{Outflow dynamics}

 The above equations for the spectral characteristics show that their evolution and, thus,
the resulting flux decay index $\alpha$, is set by the forward-shock dynamics: $\Gamma (t)$, 
$M(t)$, and $r(t)$. The energy-per-particle behind that shock is $\epsilon/n' = \Gamma$, hence 
the lab-frame energy of the forward-shock is $\Gamma^2 M c^2$ and its dynamics is given by 
$\Gamma^2 M =  E/c^2 = (M_0\Gamma_0) (t/t_0)^e$, where $M_0 \Gamma_0$ is the kinetic energy of 
the ejecta (all transferred to the forward-shock) at the time $t_0$ when the assumed power-law
energy injection ($E \propto t^e$) begins. 
The other equations for the forward-shock dynamics are $dM = Ar^{-s} (r\theta)^2 dr$, with 
$\rho = A r^{-s}$ the density of the ambient medium at radius $r$ ($s=0$ for a homogeneous medium, 
$s=2$ for a wind) and $d\theta = c_\s dt'/r = dr/(\sqrt{3} r \Gamma)$ the increase of the jet 
opening at the comoving-frame sound speed ($c_\s = c/\sqrt{3}$), and $dt = dr/(c\Gamma^2)$ 
for the observer-frame arrival time $t$ of photons emitted along the direction toward the observer. 

 Defining $f = M/M_0$, $x = r/\r$ with $\r \equiv (6M_0\Gamma_0/\pi A)^{1/(3-s)}$ 
(the radius where the lateral spreading almost doubles the initial opening of an adiabatic jet 
interacting with a homogeneous medium), and $\tau = t/t_0$, the above equations for the outflow 
dynamics become:
\begin{equation}
 \Gamma^2 = \frac{\Gamma_0}{f}\tau^e 
\label{dyn1}
\end{equation}
\begin{equation}
 \frac{\d f}{\d x}=6\Gamma_0 x^{2-s} \theta^2
\end{equation}
\begin{equation}
 \frac{\d \theta}{\d x}=\frac{1}{\sqrt{3}x\Gamma} 
\end{equation}
\begin{equation}
 \frac{\d \tau}{\d x}=\frac{\r}{2ct_0\Gamma^2}
\label{dyn2}
\end{equation}
Equation (\ref{dyn2}) allows the conversion of the jet dynamics [$\Gamma(r)$, $M(r)$, 
$\theta(r)$] from lab frame to observer frame [$\Gamma(t)$, $M(t)$, $\theta(t)$].

  For a spherical outflow, a conical jet, or a spreading jet before the jet-break time, 
$\theta$ is constant and the above equations can be solved easily, leading to
\begin{equation}
 \Gamma \propto t^{-(3-s-e)/(8-2s)} 
\label{Gm}
\end{equation}
\begin{equation}
  M \propto t^{(3-s)(1+e)/(4-s)}  (\propto nr^3)
\end{equation}
\begin{equation}
  r \propto t^{(1+e)/(4-s)} 
\end{equation}
\begin{equation}
  t' \propto t^{(5-s+e)/(8-2s)} (\propto r/\Gamma)
\end{equation}
which can be replaced in the equations for the synchrotron and inverse-Compton spectral 
characteristics (equations \ref{spsy1}--\ref{spsy3} and \ref{spic1}--\ref{spic3}) to obtain 
their evolutions and, then, the power-law index of the afterglow flux decays given in lines 
1, 2, 4, and 5 of Table \ref{spek} and Table \ref{alfa}, respectively.

\begin{table*} 
 \caption{
  Evolution of the spectral characteristics ($F_\p$ = spectral peak flux, $\nu_\i$ = energy at which 
  radiate the electrons with typical post-shock energy, $\nu_\c$ = cooling frequency) with observer 
  time $t$ for the ({\bf syn}chrotron and {\bf i}nverse-{\bf C}ompton spectra, for two types of 
  ambient media (homogeneous and $r^{-2}$ wind), for three models for the forward-shock dynamics 
  (`spherical' = wide jet, with a jet-break after last observation, `conical jet' = confined jet, 
  observed after jet-break, `spreading jet'= laterally spreading jet, observed after jet-break), 
  and a forward-shock energy increasing as $E \propto t^e$. }
\begin{tabular}{lcccccccc}
   \hline \hline
 {\sc Model} & \multicolumn{3}{c}{\sc Homogeneous medium}  & \multicolumn{3}{c}{\sc Wind-like medium} \\
             & $d\ln F_\p/d\ln t$ & $d\ln \nu_\i/d\ln t$ & $d\ln \nu_\c/d\ln t$  
             &  $d\ln F_\p/d\ln t$ & $d\ln \nu_\i/d\ln t$ & $d\ln \nu_\c/d\ln t$ \\ 
   \hline
 syn -- spherical   &  $e$   &-\fr2$(3-e)$ & -\fr2$(1+e)$ & -\fr2$(1-e)$ & -\fr2$(3-e)$ & \fr2$(1+e)$\\  \\
 syn -- conical jet & \f4$(5e-3)$ & -\fr2$(3-e)$ & -\fr2$(1+e)$ & $-(1-e)$ & -\fr2$(3-e)$& \fr2$(1+e)$\\ \\
 syn -- spreading jet & $\frac{4}{3}e-1$ & -$\frac{2}{3}(3-e)$ & $-\frac{2}{3}e$ &$-(1-e)$&$-(2-e)$&$e$   \\
 \hline
 iC -- spherical     & \f4$(5e+1)$& $-\frac{3}{4}(3-e)$ & -\f4$(5e+1)$& -1 &$-(2-e)$&$2+e$\\ \\
 iC -- conical jet   & \fr2$(3e-1)$& $-\frac{3}{4}(3-e)$ & -\f4$(5e+1)$& -\fr2$(3-e)$&$-(2-e)$&$2+e$\\ \\
 iC -- spreading jet & $\frac{4}{3}e$ &$-(3-e)$& $-(\frac{5}{3}e-1)$&$-e$&$-(3-2e)$&$1+2e$ \\
 \hline \hline
\end{tabular}
\label{spek}
\end{table*}

\subsection{Spreading jets}

\subsubsection{Adiabatic jet, homogeneous medium}
\label{adb}

 For a jet spreading laterally, the dynamics equations can be solved analytically in the case 
of no energy injection and for a homogeneous medium. That has been done by Rhoads (1999), who 
obtained an exponential jet deceleration for a lateral spreading given by 
$\theta = c_\s t'/r$: $\Gamma \propto \exp(-kr)$, $\theta \propto \exp(kr)/r$, $f \propto \exp(2kr)$
at $r > k^{-1}$, where $k = 1/\r$. Because the differential lateral spreading is 
``on top" of a conical jet and not ``over" a cylindrical jet (as in Rhoads' formalism), 
the correct lateral expansion is $d\theta = c_\s dt'/r$. For $e=0$ and $s=0$, equations
(\ref{dyn1})--(\ref{dyn2}) (which use the latter prescription for jet spreading) have the 
solution $\Gamma \propto \exp(-kr)/r$, $\theta \propto \exp(kr)$, $f \propto r^2 \exp(2kr)$ for
$r > \r$, which can be shown to lead to the same flux power-law decay indices 
as for Rhoads' prescription because the increase of the jet radius is only logarithmic 
in observer time (note that, for a nearly constant jet radius $r$, $d\theta = c_\s dt'/r$ 
implies $\theta = c_\s t'/r$). 

 The above exponential solutions for the dynamics of an adiabatic jet interacting with
a homogeneous medium can be expressed in observer time by integrating $t = t(\r) + \int_{\r}^r 
dr/(2c\Gamma^2)$, which leads to $t \simeq \r/(4c\Gamma^2)$ for $r \gg \r$. Thus
\begin{equation}
 \Gamma = \frac{1}{2} \left( \frac{\r}{ct} \right)^{1/2} \propto 
   \left( \frac{E}{n} \right)^{1/6} t^{-1/2} 
\label{gm}
\end{equation}
from where it can be shown that
\begin{equation}
 M =  \frac{4Et}{\r c} \propto E^{2/3} n^{1/3} t 
\label{m}
\end{equation}
\begin{equation}
 \theta = \frac{2}{r} \left(\frac{1}{3} \r ct \right)^{1/2} 
   \propto \left( \frac{E}{n} \right)^{1/6} \frac{t^{1/2}}{r} 
\label{teta}
\end{equation}
\begin{equation}
 t' \simeq \frac{\r}{c\Gamma} = 2 \left( \frac{\r t}{c} \right)^{1/2} \propto 
    \left( \frac{E}{n} \right)^{1/6} t^{1/2}
\label{tprim}
\end{equation}
where $E$ is the jet energy and $\r =(6E/\pi n m_\p c^2)^{1/3}$. 

\begin{table*} 
 \caption{
  Afterglow flux power-law decay index $\alpha$ ($F_\n \propto t^{-\alpha}$) for the models
  listed in Table \ref{spek}. $p$ is the power-law index of the electron distribution with energy
  behind the forward-shock: $dN/d\epsilon \propto \epsilon^{-p}$, determined from the spectral
  slope $\beta$: $p=2\beta+1$ for $\nu_\i<\nu<\nu_\c$ and $p=2\beta$ for $\nu_\i,\nu_\c<\nu$.  } 
\begin{tabular}{lcccccccc}
   \hline \hline
 {\sc Model} & \multicolumn{2}{c}{\sc Homogeneous medium}  & \multicolumn{2}{c}{\sc Wind-like medium} \\
         &  $\nu_\i<\nu<\nu_\c$ & $\nu_\i,\nu_\c<\nu$ & $\nu_\i<\nu<\nu_\c$ & $\nu_\i,\nu_\c<\nu$ \\ 
   \hline
 syn -- spherical     & 
             \f4$(3p-3)-$ \f4$(p+3)e$& \f4$(3p-2)-$ \f4$(p+2)e$& 
             \f4$(3p-1)-$ \f4$(p+1)e$& \f4$(3p-2)-$ \f4$(p+2)e$\\ \\
 syn -- conical jet   & 
             \f4$3p-$ \f4$(p+4)e$    & \f4$(3p+1)-$ \f4$(p+3)e$& 
             \f4$(3p+1)-$ \f4$(p+3)e$& \f4$3p-$ \f4$(p+4)e$   \\ \\
 syn -- spreading jet & 
            $p-$ \frc3$(p+3)e$&$p-$ \frc3$(p+2)e$   & 
            $p-$ \fr2$(p+1)e$ &$p-$ \fr2$(p+2)e$   \\ 
 \hline
 iC -- spherical     & 
             \fk8$(9p-11)-$ \fk8$(3p+7)e$& \fk8$(9p-10)-$ \fk8$(3p+2)e$& 
            $p-$ \fr2$(p-1)e$           &$p - 1-$ \fr2$pe$             \\ \\
 iC -- conical jet     & 
             \fk8$(9p-5)-$ \fk8$(3p+9)e$& \fk8$(9p-4)-$ \fk8$(3p+4)e$& 
            $p+$ \fr2 $- $\fr2$pe$        &$p-$ \fr2 $-$ \fr2$(p+1)e$     \\ \\
 iC -- spreading jet & 
             \fr2$(3p-3)-$ $\frac{1}{6}(3p+5)e$& \fr2$(3p-4)-$ \fr2$pe$ & 
             \fr2$(3p-3)-(p-2)e$              & \fr2$(3p-4) - (p-1)e$ \\
 \hline \hline
\end{tabular}
\label{alfa}
\end{table*}

\subsubsection{Energized jets}

 In the more general case of energy injection and/or for a wind-like medium, the solution to the
dynamics equations (\ref{dyn1})--(\ref{dyn2}) cannot be found analytically. They can be cast as 
logarithmic derivatives
\begin{equation}
 \frac{\d \ln f}{\d \ln \tau} \propto \frac{\theta^2}{f^2} x^{2-s} \tau^{e+1} 
\label{f}
\end{equation}
\begin{equation}
 \frac{\d \ln \theta}{\d \ln \tau} \propto \frac{\tau^{1+e/2}}{\sqrt{f} \theta x} 
\end{equation}
\begin{equation}
 \frac{\d \ln x}{\d \ln \tau} \propto \frac{\tau^{e+1}}{fx}           
\label{x}
\end{equation}
and integrated numerically, which showed that the left-hand sides are (asymptotically) 
constant, i.e. the dynamics [$M(t)$, $\theta(t)$, $r(t)$] of a spreading jet undergoing 
energy injection according to $E \propto t^e$ is a power-law in observer time. That dynamics 
is between the extreme cases of an adiabatic jet interacting with a homogeneous medium 
(which yields the fastest deceleration) and a spherical outflow with energy injection (which 
yields the slowest deceleration). Given that, in both of those two extreme cases, the dynamical 
quantities have a power-law evolution with time, it is not surprising that an intermediate 
case manifests that feature too. 

 Replacing power-law solutions for $f$, $\theta$, and $x$ in equations (\ref{f})--(\ref{x}), 
allows the calculation of the temporal exponents by requiring that the right-hand sides are 
also constants, i.e. without a $\tau$ dependence. The result is 
\begin{equation}
 M \propto t^{e+1-\frac{e}{3-s}} (\propto n r^3\theta^2) 
\label{dyn3}
\end{equation}
\begin{equation}
 \theta \propto t^{\frac{1}{2}(1-\frac{e}{3-s})} 
\end{equation}
\begin{equation}
 r \propto t^{\frac{e}{3-s}} 
\label{dyn4}
\end{equation}
 It is worth mentioning that the above dynamics of spreading jets undergoing energy injection 
can be recovered from the dynamics of a spreading adiabatic jet interacting with a homogeneous 
medium (equations \ref{m}, \ref{teta}) if one replaces $E \propto t^e$ (the prescription for
energy injection) and $r \propto \Gamma^2 t$. The latter relation is trivial for an outflow 
decelerating as a power-law with radius (as for a spherical outflow or a conical jet) and is 
approximately satisfied for an exponential deceleration (as for an adiabatic spreading jet 
interacting with a homogeneous medium), which suggests that it should also stand in any
other case of an intermediate-strength deceleration (that is verified with the numerical 
integration of the jet dynamics). Either from equations (\ref{dyn3}) and (\ref{dyn4})or by 
replacing $E$ and $n \propto r^{-s}$ in equations (\ref{gm}) and (\ref{tprim}) with the above 
prescriptions, it can be shown that
\begin{equation}
 \Gamma \propto t^{-\frac{1}{2}(1-\frac{e}{3-s})} 
\label{dyn5}
\end{equation}
\begin{equation}
  t' \propto t^{\frac{1}{2}(1+\frac{e}{3-s})} (\propto r/\Gamma)
\label{dyn6}
\end{equation}
for a spreading jet and with energy injection.

 By substituting equations (\ref{dyn3})--(\ref{dyn6}) in the relations for the synchrotron and 
inverse-Compton spectral characteristics (equations \ref{spsy1}--\ref{spsy3} and \ref{spic1}--\ref{spic3}), 
one can derive the evolutions from lines 3 and 6 of Table \ref{spek} and the power-law indices 
of the afterglow flux decay from lines 3 and 6 of Table \ref{alfa}.

  We acknowledge that the dynamics of spreading
jets undergoing energy injection was derived using a simplistic toy model for the jet lateral 
spreading, where the jet was assumed to have a uniform kinetic energy per solid angle $dE/d\Omega$, 
although both lateral spreading and energy injection could/should introduce a substantial 
angular gradient of $dE/d\Omega$, decreasing with angle measured from the jet's spine.

\section{Energy injection in Swift X-ray afterglows}
\label{swift}

\begin{table*}
 \caption{ Light-curve breaks of well-sampled Swift X-ray afterglows. GRBs in bold-face
      have afterglow light-curve breaks that can be accounted for by an adiabatic jet-break 
     (without energy injection). The GRBs listed in italics have a light-curve break that
     cannot be explained with a jet-break undergoing an energy injection continuous across 
    the break}
\begin{tabular}{cccccccccc}
   \hline \hline
   GRB & $\beta_\x$ & $t_\0$ & $\alpha_{\rm x1}$ & $t_\b$ & $\alpha_{\rm x2}$ & $t_\ended$ & 
   $E_{\rm \gamma}$ & $z$  \\
            &            & (ks) &            & (ks) &            &  (Ms)  & (erg) &      \\
            &   (1)      & (2)  &     (3)    & (4)  &     (5)    &  (6)   &  (7)  & (8) \\
   \hline 
    050315  &  0.99(.09) &  4   &  0.67(.03) &  200 &  1.33(.16) &  1   &       &      \\ 
\bf{050401} &  0.79(.13) &  0.1 &  0.77(.12) &  10  &  1.36(.08) &  1   & 4\z53 & 2.90 \\
    050712  &  1.31(.14) &  4   &  0.66(.10) &  100 &  1.29(.18) &  1.5 &       &      \\ 
    050713B &  1.08(.14) &  4   &  0.56(.05) &  70  &  1.20(.08) &  1.5 &       &      \\ 
\bf{050802} &  0.86(.06) &  0.4 &  0.71(.07) &  10  &  1.66(.08) &  1   & 4\y52 & 1.71 \\
    050803  &  1.12(.10) &  0.5 &  0.40(.05) &  10  &  1.75(.08) &  1   &       &      \\ 
\bf{050814} &  0.97(.13) &  2   &  0.82(.05) &  100 &  1.89(.18) &  1   & 2\z53 & 5.3  \\
    050822  &  1.11(.14) &  7   &  0.46(.05) &  70  &  1.09(.05) &  4   &       &      \\  
    050922C &  1.21(.11) &  0.1 &  0.92(.10) &  5   &  1.38(.10) &  0.1 & 1\z53 & 2.20 \\
\bf{051016B}&  0.89(.12) &  4   &  0.75(.08) &  40  &  1.25(.10) &  0.7 &       &      \\ 
\bf{051022} &  1.07(.12) &  10  &  1.44(.10) &  200 &  2.19(.28) &  1   &       &      \\ 
\bf{051109} &  1.06(.08) &  4   &  1.12(.05) &  70  &  1.31(.05) &  2   &       &      \\  
    060105  &  1.05(.08) &  0.1 &  0.84(.02) &  2   &  2.25(.07) & 0.03 &       &      \\ 
    060111B &  1.24(.17) &  0.2 &  0.84(.07) &  10  &  1.38(.16) &  0.2 &       &      \\ 
    060115  &  1.10(.12) &  1   &  0.56(.15) &  30  &  1.31(.17) &  0.4 & 8\y52 & 3.53 \\
\bf{060204B}&  1.21(.13) &  5   &  1.17(.07) &  40  &  1.76(.24) &  0.3 &       &      \\  
    060210  &  1.08(.05) &  4   &  0.99(.03) &  100 &  1.43(.15) &  1.5 & 4\z53 & 3.91 \\
\it{060413} &  0.90(.10) &  5   &  0.08(.08) &  30  &  2.79(.16) &  0.2 &       &      \\ 
\bf{060428} &  0.94(.15) &  1   &  0.63(.05) &  100 &  1.19(.10) &  1   &       &      \\ 
\it{060510} &  0.88(.09) &  0.2 &  0.05(.10) &  5   &  1.44(.05) &  0.4 &       &      \\
    060604  &  1.17(.12) &  2   &  0.50(.08) &  30  &  1.22(.08) &  1   & 5\u51 & 2.68 \\
\it{060607} &  0.61(.06) &  0.6 &  0.40(.05) &  20  &  3.16(.12) &  0.1 &1.5\z53& 4.05 \\
\it{060614} &  0.90(.09) &  3   &  0.04(.08) &  40  &  2.17(.12) &  2   & 2\u51 & 0.13 \\
\it{060729} &  1.02(.04) &  0.6 &  0.08(.03) &  100 &  1.36(.02) &  3   & 5\u51 & 0.54 \\
\bf{060813} &  0.92(.09) &  0.2 &  0.99(.02) &  40  &  2.06(.30) &  0.2 &       &      \\
    060814  &  1.12(.07) &  1   &  0.34(.12) &  6   &  1.01(.03) &      & 7\y52 & 0.84 \\
    060814  &            &      &  1.01(.03) &  70  &  1.74(.16) &  0.6 &       &      \\
    060908  &  1.13(.19) &  0.2 &  0.81(.12) &  1   &  1.34(.10) &  0.4 & 7\y52 & 1.88 \\
\it{061021} &  0.99(.04) &  4   &  0.92(.03) &  100 &  1.16(.05) &  4   &       &      \\
\it{061121} &  0.90(.06) &  0.3 &  0.02(.10) &  10  &  1.56(.05) &  2   & 2\z53 & 1.31 \\
    061222  &  0.93(.06) &  0.3 &  0.36(.08) &  10  &  1.06(.05) &      & 3\z53 & 2.09 \\
    061222  &            &      &  1.06(.03) &  100 &  1.70(.05) &  1.5 &       &      \\
\bf{070107} &  1.08(.08) &  6   &  1.00(.05) &  100 &  1.84(.16) &  0.8 &       &      \\
    070129  &  1.28(.13) &  1   &  0.19(.15) &  20  &  1.21(.07) &  2   &       &      \\
    070220  &  0.56(.12) &  0.6 &  0.76(.13) &  3   &  1.14(.05) &      &       &      \\ 
\bf{070220} &            &      &  1.14(.05) &  20  &  2.01(.13) &  0.1 &       &      \\
\it{070306} &  0.94(.08) &  0.7 &  0.05(.08) &  30  &  1.88(.07) &  1   & 8\y52 & 1.50 \\
    070328  &  1.14(.09) &  0.15&  0.38(.08) &  0.8 &  1.12(.05) &      &       &      \\
\bf{070328} &            &      &  1.12(.05) &  5   &  1.58(.03) &  1   &       &      \\
    070420  &  0.97(.09) &  0.3 &  0.27(.10) &  3   &  1.21(.08) &      &       &      \\ 
    070420  &            &      &  1.21(.08) &  50  &  1.83(.13) &  0.5 &       &      \\
    070508  &  0.82(.11) &  0.4 &  0.99(.05) &  2   &  1.48(.10) &      &       &      \\
    070508  &            &      &  1.48(.10) &  60  &  1.67(.13) &  0.6 &       &      \\
\bf{070529} &  0.98(.12) &  0.2 &  0.76(.20) &  2   &  1.31(.08) &  0.4 &1.2\z53& 2.50 \\
    070628  &  1.04(.08) &  1   &  0.37(.07) &  7   &  1.19(.05) & 0.08 &       &      \\
\bf{071020} &  0.61(.13) &  0.1 &  1.23(.03) &  80  &  1.64(.15) &  1.5 & 9\y52 & 2.15 \\
    080319B &  0.82(.06) &  0.1 &  1.47(.02) &  2   &  1.81(.02) & 0.07 &1.3\w54& 0.94 \\
\bf{080319C}&  0.61(.10) &  0.4 &  0.84(.08) &  4   &  1.53(.05) &  0.3 &1.3\z53& 1.95 \\
    080320  &  0.97(.10) &  1   &  0.66(.05) &  80  &  1.19(.07) &  2   &       &      \\
    080328  &  0.95(.09) &  0.3 &  0.62(.10) &  3   &  1.21(.05) & 0.04 &       &      \\
    080413B &  0.94(.06) &  0.1 &  0.73(.03) &  4   &  0.98(.05) &      & 2\y52 & 1.10 \\
\bf{080413B}&            &      &  0.98(.05) &  70  &  1.49(.12) &  0.6 &       &      \\
    080430  &  1.04(.06) &  0.6 &  0.45(.05) &  50  &  1.12(.07) &  3   & 5\u51 & 0.77 \\
    080703  &  0.49(.10) &  0.1 &  0.58(.05) &  20  &  2.01(.17) &  0.1 &       &      \\
\bf{080710} &  1.01(.11) &  3   &  1.05(.08) &  30  &  1.71(.15) &  0.3 &       &      \\
\bf{080721} &  0.91(.05) &  0.1 &  0.82(.02) &  2   &  1.65(.02) &  1.4 &1.2\w54& 2.59 \\
\bf{080916} &  0.89(.12) &  0.3 &  0.68(.07) &  40  &  1.24(.10) &  1.4 & 9\u51 & 0.69 \\
    081007  &  1.10(.13) &  0.5 &  0.77(.05) &  40  &  1.24(.08) &  1.5 &1.3\u51& 0.53 \\
\bf{081008} &  0.99(.07) &  0.6 &  0.88(.05) &  20  &  1.82(.23) &  0.3 & 8\y52 & 1.97 \\
    081028  &  1.04(.05) &  20  &  1.30(.10) &  80  &  2.30(.20) &  0.5 &       &      \\
\it{081029} &  0.96(.09) &  3   &  0.45(.10) &  20  &  2.18(.13) &  0.1 &       &      \\ 
 \hline \hline
\end{tabular}
\label{ag1}
\end{table*}

\setcounter{table}{2}

\begin{table*}
 \caption{ (continued) }
\begin{tabular}{ccccccccc}
   \hline \hline
   GRB & $\beta_\x$ & $t_\0$ & $\alpha_{\rm x1}$ & $t_\b$ & $\alpha_{\rm x2}$ & $t_\ended$ &
   $E_{\rm \gamma}$ & z  \\ 
            &            & (ks) &            & (ks) &            &  (Ms)  & (erg) &      \\
            &   (1)      & (2)  &     (3)    & (4)  &     (5)    &  (6)   &  (7)  & (8 ) \\
   \hline 
\bf{081203} &  1.04(.09) &  0.2 &  1.13(.03) &  7   &  1.85(.08) &  0.3 & 3\z53 & 2.05 \\
    081221  &  1.50(.12) &  0.4 &  0.60(.08) &  1   &  1.28(.05) &      &       &      \\
    081221  &            &      &  1.28(.05) &  100 &  1.60(.23) &  0.5 &       &      \\
    081222  &  1.00(.06) &  .08 &  0.85(.02) &  2   &  1.18(.05) &      & 3\z53 & 2.77 \\
    081222  &            &      &  1.18(.05) &  70  &  1.83(.15) &  0.7 &       &      \\
\bf{090205} &  1.02(.14) &  1   &  0.77(.07) &  20  &  1.84(.30) &  0.2 &1.5\y52& 4.65 \\
    090401B &  0.93(.10) & 0.08 &  1.14(.02) &  0.5 &  1.52(.05) &  0.8 &       &      \\ 
    090404  &  1.57(.13) &  0.7 &  0.14(.08) &  10  &  1.21(.03) &  1   &       &      \\
    090407  &  1.34(.12) &  10  &  0.61(.07) &  100 &  1.84(.18) &  1   &       &      \\
    090418  &  1.03(.09) &  0.2 &  0.28(.18) &  3   &  1.58(.05) &  0.3 &1.6\z53& 1.61 \\
\it{090424} &  0.95(.09) &  2   &  1.20(.05) &  300 &  1.44(.08) &      &       &      \\
    090424  &            &      &  0.87(.02) &  2   &  1.20(.05) &  0.3 & 4\y52 & 0.54 \\
    090510  &  0.76(.12) &  0.1 &  0.74(.05) &  1.5 &  2.20(.16) & 0.06 & 2\y52 & 0.90 \\
    090516  &  1.08(.07) &  4   &  0.88(.10) &  20  &  1.75(.07) &  0.5 &       &      \\ 
    090618  &  0.92(.05) &  0.5 &  0.61(.03) &  5   &  1.44(.02) &      & 2\z53 & 0.54 \\
    090618  &            &      &  1.44(.02) &  300 &  1.83(.13) &  3   &       &      \\
    090813  &  0.89(.09) &  0.1 &  0.18(.10) &  0.5 &  1.29(.02) &  0.7 &       &      \\
    091018  &  0.98(.08) & 0.06 &  0.41(.07) &  0.7 &  1.20(.05) &      & 5\u51 & 0.97 \\
    091018  &            &      &  1.20(.05) &  40  &  1.61(.13) &  0.5 &       &      \\
    091020  &  1.09(.06) &  0.2 &  0.80(.10) &  10  &  1.37(.03) &  1   & 8\y52 & 1.71 \\
    091029  &  1.12(.07) &  0.8 & -0.23(.20) &  4   &  0.86(.07) &      & 8\y52 & 2.75 \\
    091029  &            &      &  0.86(.07) &  100 &  1.23(.12) &  2   &       &      \\  
    091127  &  0.80(.11) &  3   &  1.20(.02) &  100 &  1.60(.07) &  4   &       &      \\
    091130B &  1.37(.14) &  3   &  0.34(.07) &  50  &  1.14(.07) &  2   &       &      \\
\bf{091208B}&  1.04(.13) &  5   &  1.10(.07) &  100 &  1.57(.24) &  1   &       &      \\
\bf{100302} &  0.94(.11) &  5   &  0.66(.17) &  50  &  0.92(.13) &  1   & 3\y52 & 4.81 \\
\it{100508} &  0.43(.11) &  1   &  0.39(.07) &  20  &  2.32(.15) &  0.2 &       &      \\
    100522  &  1.28(.13) &  0.2 &  0.45(.12) &  3   &  0.84(.08) &      &       &      \\  
    100522  &            &      &  0.84(.08) &  40  &  1.37(.15) &  0.4 &       &      \\
    100619  &  1.29(.12) &  0.3 &  0.86(.03) &  30  &  1.22(.13) &  0.8 &       &      \\
    100621  &  1.31(.11) &  0.5 &  0.56(.08) &  7   &  0.99(.05) &      & 4\y52 & 0.54 \\
    100621  &            &      &  0.99(.05) &  100 &  1.61(.12) &  2   &       &      \\
    100704  &  1.12(.10) &  0.6 &  0.98(.02) &  300 &  1.84(.18) &  1   &       &      \\
    100728  &  0.90(.07) &  1   &  1.21(.01) &  30  &  1.54(.08) &  0.6 &       &      \\  
\it{100814} &  0.89(.04) &  4   &  0.52(.03) &  100 &  1.94(.07) &  2   &1.6\z53& 1.44 \\
    100906  &  1.04(.08) &  0.3 &  0.53(.16) &  6   &  1.69(.07) &      & 2\z53 & 1.73 \\
\bf{100906} &            &      &  1.69(.07) &  60  &  2.21(.18) &  0.3 &       &      \\
\bf{101023} &  1.14(.10) &  3   &  1.10(.07) &  30  &  1.43(.10) &  0.6 &       &      \\
\it{101024} &  0.82(.12) &  0.1 &  0.00(.16) &  1   &  1.37(.07) &  0.1 &       &      \\
    101117B &  1.17(.16) & 0.06 &  0.73(.08) &  2   &  1.17(.10) &  0.1 &       &      \\
    110102  &  1.13(.06) &  0.5 &  0.31(.25) &  5   &  1.38(.05) &      &       &      \\  
\bf{110102} &            &      &  1.38(.05) &  100 &  2.05(.26) &  1   &       &      \\
\bf{110213} &  0.96(.06) &  1   &  0.97(.07) &  10  &  1.99(.08) &  0.6 &       &      \\
    110223B &  0.73(.15) &  0.1 &  0.86(.05) &  5   &  1.16(.08) &  0.4 &       &      \\
    110420  &  1.05(.08) &  0.2 &  0.23(.08) &  3   &  1.18(.03) &      &       &      \\
\bf{110420} &            &      &  1.18(.03) &  200 &  2.09(.25) &  2   &       &      \\  
    110422  &  0.89(.07) &  0.8 &  1.01(.03) &  8   &  1.52(.03) &  0.7 &       &      \\
\it{110503} &  0.91(.05) &  0.2 &  1.06(.02) &  20  &  1.39(.05) &  0.8 &1.7\z53& 1.61 \\
    110709  &  1.06(.11) &  0.2 &  0.78(.03) &  7   &  1.70(.07) &  0.3 &       &      \\
    110709B &  1.13(.06) &  2   &  1.01(.03) &  100 &  1.57(.07) &  1   &       &      \\
\bf{110715} &  0.86(.10) & 0.08 &  0.57(.05) &  1   &  1.52(.10) & 0.03 & 4\y52 & 0.82 \\
    110731  &  0.85(.10) & 0.08 &  1.16(.03) &  30  &  1.29(.10) &  0.8 & 4\z53 & 2.83 \\
    110915  &  1.25(.14) &  0.3 &  0.88(.05) &  10  &  1.44(.05) &  0.4 &       &      \\
    111008  &  0.96(.07) &  0.3 &  0.07(.18) &  3   &  1.12(.07) &      & 4\z53 & 4.99 \\
    111008  &            &      &  1.12(.07) &  50  &  1.54(.12) &  0.5 &       &      \\  
    111228  &  1.05(.07) &  0.5 &  0.25(.07) &  10  &  1.13(.03) &  1   & 3\y52 & 0.71 \\
 \hline \hline
\end{tabular}
\begin{minipage}{14cm} 
  \vspace*{3mm}
 (1): X-ray spectral slope (90 percent CL) \\
 (2): epoch of first Swift X-ray measurement \\
 (3): index of X-ray flux power-law decay at $t_\0 - t_\b$ (90 percent CL) \\
 (4): epoch of X-ray light-curve break \\
 (5): index of X-ray flux power-law decay at $t_\b - t_\ended$ (90 percent CL) \\
 (6): epoch of last measurement  \\
 (7): 10 keV--10 MeV isotropic energy release during prompt phase (uncertain by up to 50 percent)  \\
 (8): redshift \\
\end{minipage}
\end{table*}

\subsection{Sample selection}

 Out of the 634 afterglow light-curves listed at the Swift-XRT light-curve repository (Evans et 
al 2007) posted until the end of year 2011, we have retained only those for which 
$i)$ a light-curve break is obvious or seems possible, 
$ii)$ a power-law light-curve segment is monitored for longer than a factor 5 in time before and 
after the break, and 
$iii)$ the spectral slope $\beta_\x$ listed at the Swift-XRT spectra repository (Evans et al 2007) 
has a 90 percent confidence level (CL) less than 0.16.

 The light-curves have been piece-wise fit with power-laws, using measurements sufficiently far 
from the break time, to capture the asymptotic power-law flux decay. Shallow breaks for which 
the data could be fit satisfactorily (i.e. reduced $\chi^2_\n \simeq 1$) with a single power-law 
were excluded. The 90 percent confidence level ($\sigma (\alpha_\x)$) was determined from 
the variation of $\chi^2$ around the best power-law fit. 

 As shown in Table \ref{alfa}, for an adiabatic outflow ($e=0$), the power-law index is $\alpha_\x = 
3p/4$ + const = $3\beta_\x$/2 + const, with the coefficient of $\beta_\x$ being as large as 9/8 
for other models. Thus, an equal importance of measurement uncertainties $\sigma (\alpha_\x)$ and 
$\sigma (\beta_\x) \leq 0.16$ requires that we retain only afterglows whose decay indices can be 
determined with a 90 percent CL smaller than $1.5 \sigma (\beta_\x) \simeq 0.24$. 

 The above two basic selection criteria led to the sample of 98 afterglows and 117 X-ray light-curve
breaks listed in Table \ref{ag1}. Forty-four afterglows have temporal coverage starting
from the burst end (for them, we determine the fractional increase of the outflow energy during 
the afterglow phase) and a measured redshift, which allows the calculation of the GRB output and
of the afterglow jet energy. 

 The light-curves of the afterglows listed in Table \ref{ag1} are morphologically
diverse. A minority display a continuous, slowly-decaying power-law flux from the burst phase to
the afterglow, the afterglow beginning being chosen when the flux fluctuations disappear.
The majority of GRBs display a sharp decay after the prompt phase (the GRB tail) followed
by a much slower decay and spectral hardening; the afterglow beginning is chosen at the epoch
when the slower flux decay starts. 


 The closure relations $\alpha-\beta$ listed in Table \ref{alfa} are used to calculate the 
required energy injection indices from the measured pre- and post-break X-ray flux decay 
indices $\alpha$ and the X-ray slope $\beta$ (which gives the electron index $p$). 
For a model to be validated, the resulting energy injection index $e$ 
$i)$ should be positive (i.e. $e + \sigma_e > 0$, with the uncertainty $\sigma_e$ calculated 
   from the propagation of $\sigma(\alpha_\x)$ and $\sigma(\beta_\x)$), and
$ii)$ for jet-break models, should lead to a decelerating outflow, so that the jet boundary
 can become visible to the observer at some time. From equation \ref{Gm}, this condition
 is $e < 3-s$.

\subsection{Adiabatic jets}
\label{adiabatic}

 For adiabatic jets ($e=0$), the observables $\alpha_{\rm x1}$ and $\alpha_{\rm x2}$ yield two constraints.
As there are no free model parameters, aside the various possibilities related to the jet
dynamics, location of the cooling frequency, ambient medium type, and dominant X-ray emission
process, the model is over-constrained. Thus, it should not be surprising that only a small fraction 
of Swift X-ray breaks satisfy $|e| < \sigma_e$ both before and after the break, i.e. can be 
explained by any single variant of the adiabatic jet break model (upper part of Table \ref{mod}).

\begin{table}
 \caption{ Number of breaks that can be accommodated by an adiabatic jet-break, a jet-break with 
  steady energy injection, or by an energy-injection break, out of the 117 breaks listed in Table
  \ref{ag1}. Models are identified by the origin
  of the X-ray light-curve break (``J" = jet-break, ``EI" = energy-injection break), radiation
  process dominant in the X-rays (``syn" = synchrotron, ``iC" = inverse-Compton) and by
  the jet dynamics (``cjet" = conical jet, ``sjet" = spreading jet, ``sph" = spherical -- 
  see Table \ref{alfa}).  }
\begin{tabular}{lcccccccc}
   \hline \hline
 {\sc model} &  \multicolumn{2}{c}{\sc Homog medium}  & \multicolumn{2}{c}{\sc Wind-like medium} \\
                     &  $\nu < \nu_\c$ &  $\nu_\c < \nu$ &   $\nu < \nu_\c$ &  $\nu_\c < \nu$ \\
   \hline
 \multicolumn{5}{l}{adiabatic jet-break}    \\
   \hline
 J--syn--cjet  & 3 & 11 & 0 & 9  \\
 J--syn--sjet  & 0 & 7  & 0 & 7  \\
 J--iC--cjet  & 2 & 15 & 0 & 9  \\
 J--iC--sjet  & 2 & 7  & 0 & 7  \\
   \hline
 \multicolumn{5}{l}{jet-break with or without steady EI}    \\
  \hline
 J--syn--cjet  & 48 & 47 & 12 & 29  \\
 J--syn--sjet  & 19 & 39 & 15 & 42  \\
 J--iC--cjet  & 64 & 52 & 0  & 31  \\
 J--iC--sjet  & 67 & 18 & 0  & 16  \\
  \hline
 \multicolumn{5}{l}{energy-injection break} \\
   \hline
 EI--syn--cjet  & 111 &  98 & 113 & 77  \\ 
 EI--syn--sjet  & 115 & 106 & 115 & 105 \\
 EI--syn--sph   &  77 &  31 & 106 & 31  \\
 EI--iC--cjet  & 114 &   4 &  116 & 80  \\ 
 EI--iC--sjet  & 114 &  47 &  114 & 54  \\
 EI--iC--sph   & 105 &  37 &  115 & 36  \\ 
 \hline \hline
\end{tabular}
\label{mod}
\end{table}

 Evidently, Swift X-ray afterglows may not be all of the same forward-shock model type.
Even when we take into account all these possible variants, we find that adiabatic jet-break
model can explain only 30 afterglows (indicated in boldface in Table \ref{ag1}), 
i.e. about 30 percent of the sample considered here. Six of those 30 breaks  occur in afterglows 
with two light-curve breaks, the break compatible with an adiabatic jet-break being always the 
second. For all six, there is always a spherical energy-injection break model with $e=0$ after 
the break that accounts for the first break, and compatible with the jet-break model, i.e. the 
two break models have the same location of cooling frequency relative to the X-rays, the same 
type of external medium, and the same X-ray emission process.

\subsection{Energized jets}

\subsubsection{Jet-breaks} 
\label{jbei}

\begin{figure}
\centerline{\psfig{figure=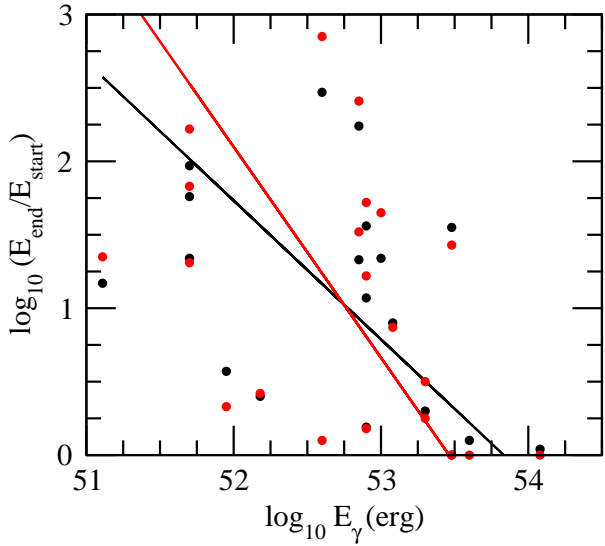,width=8cm}}
\caption{ Fractional increase of the jet energy from (source frame) 100 s to 100 ks ($E_\ended/E_\0$)
   versus GRB output at 10 keV -- 10 MeV ($E_\gamma$) and whose X-ray light-curve breaks can be 
   accommodated by a conical jet-break with steady energy injection (other model details: X-ray 
   higher than the synchrotron cooling frequency, external medium is homogeneous). Black points 
   are for a model where synchrotron is the dominant X-ray radiative process, red points for an 
   inverse-Compton model. These are the afterglows listed in Table \ref{ag1} for which a redshift is known
   and whose light-curve breaks (first break, if two exist) can be accommodated by the models described
   above. In natural scale, $E_\ended/E_\0$ and $E_\ended/E_\0$ are uncertain by up to 50 percent, hence 
   their logarithms are uncertain by up to 0.20 .}
\label{f1} 
\end{figure}

 For the jet-break model, we require that the X-ray light-curve break arises solely from the 
outflow collimation, without any contribution from a break in the injected power, i.e. we require 
a constant energy injection index $e$ across the X-ray light-curve break: $(e_\1 - e_\2)^2 < 
\sigma^2_{e_\1} + \sigma^2_{e_\2}$. Thus, the model is still over-constrained, with two observables 
($\alpha_{\rm x1}$ and $\alpha_{\rm x2}$) that must be accommodated by a single model
parameter ($e$). Considering that the jet-break model with steady energy injection injection is 
less over-constrained than the adiabatic jet model, it may not be surprising that the former
model can account for a larger number of Swift breaks (mid part of Table \ref{mod}).

 With allowance for all jet-break model variants listed in Table \ref{mod} (i.e. for either 
location of the cooling frequency relative to the X-ray, for either radiation process, for
either type of medium, and either post jet-break dynamics, conical or spreading), the jet-break 
model with steady energy injection (including adiabatic jet-breaks, for which the energy injection
parameter $e$ is consistent with zero) can account for 103 of the 117 X-ray light-curve breaks 
considered here (the unexplained breaks are shown in italics in Tables \ref{ag1}).
37 of those potential jet-breaks occur in afterglows with two light-curve breaks; with the
exception of the first break of afterglow 081222, all other 36 jet-breaks can be paired consistently 
with an energy-injection break for the 
remaining break, i.e. the two break models have the same medium stratification, the same X-ray 
radiative process, the same location of X-ray relative to the cooling frequency, and compatible 
ejecta dynamics. 

 Because only one break can arise from the outflow collimation and 19 afterglows in our sample 
have two light-curve breaks, a more correct way of assessing the success of the jet-break model 
with steady energy injection is to say that it can explain at least one of the breaks observed 
for 85 afterglows out of 98 or, conversely, that it cannot account for any of the X-ray light-curve 
breaks of 12 percent of afterglows. 
Excluding the 30 breaks (of 30 afterglows) that can be accounted for with an adiabatic jet-break,
the jet-break model with non-zero energy injection can explain one X-ray light-curve break for 55 
of the 98 afterglows considered here, which argues strongly that energy injection may be often 
present in Swift afterglows.

\subsubsection{Energy injection breaks} 
\label{eibr}

 Given that 12 percent of afterglows have an X-ray light-curve break that cannot be explained 
with a jet-break even when a steady energy injection is allowed, and that some afterglows display 
two breaks, out of which only one can be a jet-break, it is natural to speculate that some X-ray 
light-curve breaks must arise from a change in the power at which energy is added to the 
forward-shock. In the extreme, one could speculate that all X-ray breaks could be such energy 
injection breaks, but the large fraction of breaks that can be explained as jet-breaks argues 
against that extreme speculation.

 The broken power-law energy injection model is not over-constrained, as there are two observables
($\alpha_{\rm x1}$ and $\alpha_{\rm x2}$) for two model parameters ($e_\1$ and $e_\2$), the only 
restriction being that $e > 0$.  Thus, it should not be surprising that there are variants of 
the energy-injection break model that can account for most of the Swift X-ray breaks (lower part 
of Table \ref{mod}). There is only one X-ray break that cannot be explained with an energy 
injection break: GRB afterglow 060607. From Table \ref{alfa}, it can be seen that, for $\beta_\x < 
3/2$, the steepest possible decay for an adiabatic outflow is that for the inverse-Compton emission 
from a conical jet interacting with a wind, and for $\nu_\x < \nu_\c$: $\alpha_\x = p + 1/2 = 
2\beta_\x + 3/2$ \footnotemark. 
For GRB afterglow 060607, $\beta_\x = 0.61 \pm 0.06$, hence the steepest possible 
decay for the forward-shock emission has $\alpha_\x = 2.72 \pm 0.12$, which is slower than the 
measured post-break $\alpha_{\rm x2} = 3.16 \pm 0.07$. 
\footnotetext{ For $\beta_\x > 1/2$, the fastest decay $\alpha_\x = 2\beta_\x + 3/2$, exceeds 
  the $\alpha_\x = \beta_\x + 2$ limit for the large-angle emission decay, resulting when the 
  emission from a spherical outflow stops suddenly; however, for a jet seen after the jet-break 
  time, the large-angle emission does not exist because there is no radiating fluid at angles 
  larger than $\Gamma^{-1}$}

\section{Implications of Energy Injection}

\subsection{Afterglow decay rate -- GRB output correlation}
\label{fag}

 After having identified the power-law energy injection that accommodates each of the X-ray 
light-curve breaks, one can calculate the increase in the forward-shock energy from first to 
last X-ray measurement. The interesting result is that the fractional increase of the 
forward-shock energy over a source-frame time-interval common to all afterglows (from 100 s 
to 100 ks), $f_\ag \equiv E_\ended/E_\0$, is anticorrelated with the isotropic-equivalent 
GRB output $E_\rgamma$, for any model variant that can account for a significant fraction of 
the afterglow breaks listed in Table \ref{ag1}.
 
\begin{figure}
\centerline{\psfig{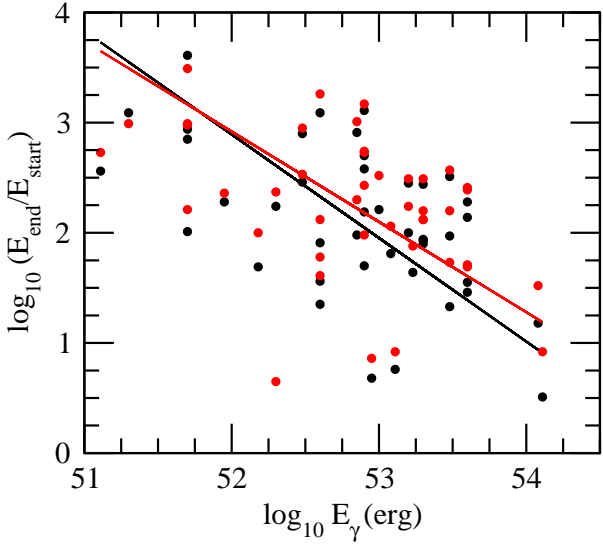}}
\caption{ Similar to Fig \ref{f1}, but for energy-injection break as the origin of the X-ray
    light-curve break, with the afterglow outflow being a conical jet and the jet-break 
    having occurred at early times (before the first X-ray measurement). The ambient medium
    is homogeneous and X-ray assumed to be below the cooling frequency. Black points are 
    for synchrotron as the dominant X-ray process, red points are for an inverse-Compton model.}
\label{f2} 
\end{figure}

 The $\Egfag$ anticorrelation is illustrated in Figure \ref{f1} for two jet-break models and 
in Figure \ref{f2} for two energy-injection break models. 
For the synchrotron jet-break model (black points in Figure \ref{f1}), the linear correlation 
coefficient in log-log space is $r = -0.42 \pm 0.08$, yielding a probability for a stronger 
anticorrelation $prob < 0.080$  in the null hypothesis.
For the inverse-Compton jet-break model (red points in Figure \ref{f1}), $r = -0.40 \pm 0.07$ 
and $prob < 0.076$.  
For the synchrotron energy-injection break model (black points in Figure \ref{f2}), $r = -0.54 
\pm 0.04$, and $prob < 0.52 \times 10^{-3}$, while for the inverse-Compton energy-injection
break model (red points in Figure \ref{f2}), $r = -0.43 \pm 0.06$, and $prob < 0.70 \times 10^{-2}$.
Similarly high chance probabilities are obtained for other models; thus the $\Egfag$ anticorrelation
is only ``tentative".

 The above linear correlation coefficients (Pearson's) $r$ have been calculated assuming a 50 
percent uncertainty (in natural scale) for both quantities plotted, the chance probabilities 
of those anticorrelations have been estimated conservatively for a linear correlation coefficient 
$|r| - \sigma(r)$. The best-fits shown in Figures \ref{f1} and \ref{f2} were obtained by minimizing 
$\chi^2$ using the uncertainties of both quantities. 

\begin{figure}
\centerline{\psfig{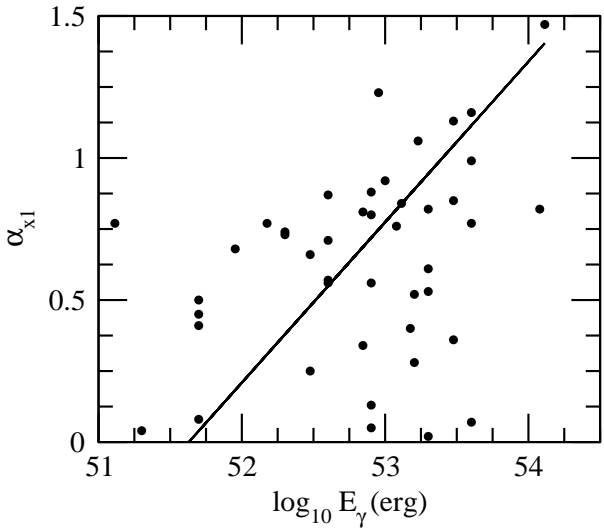}}
\caption{ Correlation of power-law index $\alpha_{\rm x1}$ of the early afterglow X-ray flux decay 
   and GRB output $E_\rgamma$ for 44 Swift afterglows. Linear correlation coefficient is 
   $r = 0.34 \pm 0.05$, with a chance probability $prob < 0.026$ . }
\label{f3}
\end{figure}

 That $i$) the fractional increase of the forward-shock energy over a fixed time interval is 
determined by the exponent $e$ of the power-law energy injection which accommodates the 
measured X-ray flux decay index $\alpha_\x$, and that $ii$) $d\alpha/de < 0$ (a stronger energy 
injection yields a slower flux decay) suggest that the $\Egfag$ correlation may originate from 
a $\log E_\rgamma - \alpha_\x$ correlation. Indeed, as illustrated in Figure \ref{f3}, more energetic 
bursts are followed by faster decaying X-ray afterglows. 
The correlation between these two observables is slightly stronger than the $\Egfag$ correlation
for afterglows with jet-breaks at the X-ray light-curve epoch, but significantly weaker if the 
X-ray light-curve breaks are energy-injection breaks.

 The $\Egfag$ correlation implies that either the efficiency of the GRB mechanism or the energy
of the ejecta producing the GRB emission are anticorrelated with the kinetic energy of the ejecta
interacting with the forward-shock after the burst phase.

 If the dissipation mechanism leading to the production of $\gamma$-rays is internal shocks
(Rees \& Meszaros 1994) in a fluctuating relativistic wind, then the burst dissipation efficiency
is determined by the ratio between the Lorentz factors of interacting shells.
For an {\sl impulsive} ejecta release (short-lived engine), all the ejecta are produced quasi-instantaneously
and the energy injection occurs as the inner parts of the outflow catch-up with the ever-decelerating
forward-shock (Rees \& Meszaros 1998). In this model, the $\Egfag$ correlation implies an 
anticorrelation between the average shells' Lorentz factor ratio with the energy of the shells 
behind the leading edge of the outflow, smoother outflows (i.e. with a smaller Lorentz factor 
contrast) having more energy localized toward the trailing edge. 
 For an {\sl extended} ejecta release (long-lived engine), where the ejecta are released over a 
source-frame duration comparable to the observer-frame time during which energy injection takes place, 
the $\Egfag$ correlation would imply an anticorrelation among bursts between the Lorentz factor 
contrast in the leading, GRB-producing part of the outflow and the kinetic energy of the outflow 
expelled by the central engine at later times. This is a puzzling implication, which makes the 
long-lived engine scenario less likely.

 Irrespective of the dissipation mechanism, if the GRB output is correlated with the energy of
the GRB-producing ejecta, then the $\Egfag$ correlation implies a relationship between the GRB
ejecta energy and the radial distribution of the afterglow outflow energy. For a short-lived
central engine, where the GRB and afterglow outflows are the same, the $\Egfag$ correlation would
require that more energetic outflows have less energy behind the leading front. For a long-lived
central engine, where the GRB outflow is the early afterglow blast-wave, the $\Egfag$ correlation
implies that engines that are initially more energetic release less energy after the burst. Perhaps
this scenario appears more plausible because it opens the possibility of a constant total ejecta energy.
Unfortunately, we do not find that $f_\ag E_\rgamma$ is universal; instead, its distribution extends
over at least two decades, for the $f_\ag$ calculated in any light-curve break model.

 We note that the detection of X-ray flares in some Swift afterglows indicates an extended ejecta
release, but the energetic output of those flares is only about 10 percent of the prompt GRB output
(Falcone et al 2007). In contrast, the energy injection required during the afterglow phase corresponds
to an increase of the outflow energy after the burst by a factor up to $10^4$ (Figures \ref{f1} and \ref{f2}).
To reconcile the flare and afterglow energetics, the internal shocks that produce flares must be very
inefficient X-ray radiators compared to the forward-shock (see also Maxham \& Zhang 2009) or, else, 
the X-ray flares could be produced by the central engine, bearing little correlation with the energy
of the ejecta released after the burst.

\subsection{Afterglow energetics}
\label{jeten}

 The large increase of the forward-shock energy, by a factor $f_\ag=10-300$ (Fig \ref{f1}) or 
$f_\ag=10-10^4$ (Fig \ref{f2}), required to account for the light-curve breaks of Swift X-ray 
afterglows raises the question if the total outflow energy is not too large and incompatible 
with the expected energy reservoir for a few solar-masses black-hole and its sub solar-mass 
torus formed in the collapse of a WR massive star. 

 The jet energy can be estimated by assuming that, at the first X-ray measurement ($t_\0$), 
the isotropic-equivalent forward-shock energy is the GRB output $E_\rgamma$. This assumption 
introduces an underestimation of the jet energy by a factor up to a few because energy could 
be injected into the forward-shock even before $t_\0$ (100 -- 1000 s), maybe starting as early 
as the burst end (10 -- 100 s), at a rate $E \propto t^{1/3}-t^1$ (this is the range of energy
injection laws that accommodate the early X-ray flux decays of Table \ref{ag1}), 
and with the estimated jet energy satisfying $E_\j \propto [E(t_\0)]^{3/4}$ for a homogeneous 
medium and $E_\j \propto [E(t_\0)]^{1/2}$ for a wind. 
For a short-lived engine, where all the outflow produces the burst emission and only its leading 
edge yields the early afterglow, we have $E(t_\0) \ll E_\rgamma$; in this case, the total jet 
energies calculated here may be substantially overestimated.
For a long-lived engine, where the leading part of the outflow produces both the burst and early 
afterglow emissions, $E(t_\0) = E_\rgamma$ is equivalent to assuming a (reasonable) 50 percent GRB 
efficiency. 

 The jet initial aperture $\theta_\j$, needed to calculate the jet energy from isotropic equivalents, 
can be constrained as following. 
For the model of a narrow jet whose edge becomes visible before the first X-ray measurement 
and the X-ray light-curve break is an energy injection, $\theta_\j < 1/\Gamma(t_\0)$.
For the model where the X-ray light-curve break (at $t_\b$) is a jet-break undergoing a steady energy 
injection, $\theta_\j = 1/\Gamma(t_\b)$.
For the model of a wide jet whose edge becomes visible after the last X-ray measurement ($t_\ended$),
$\theta_\j > 1/\Gamma(t_\ended)$. 
Prior to the jet-break, the jet Lorentz factor is given by the dynamics of a spherical blast-wave:
\begin{equation}
 \Gamma (t) = 800\, \left[ \frac{E_{53}(t)}{n_\o} \right]^{1/8} \left( \frac{t}{z+1}\right)^{-3/8}
\end{equation}
for a homogeneous medium of particle density $n_\o$ per $\rm{cm^{-3}}$ and
\begin{equation}
 \Gamma (t) = 175\, \left[ \frac{E_{53}(t)}{A_*} \right]^{1/4} \left( \frac{t}{z+1}\right)^{-1/4}
\end{equation}
for a wind-like medium of density $n \propto A r^{-2}$ normalized to that corresponding to progenitor 
mass-loss rate-to-terminal 
speed ratio of $(dM/dt)/v = 10^{-5}\, (M_\odot \, yr^{-1})/(10^3\; {\rm km\, s^{-1}})$, respectively,
with $E_{53}(t) = 10^{-53} E(t)$ being the blast-wave isotropic-equivalent energy at observer
time $t$. The latter is calculated assuming that $E(t_\0) = E_\rgamma$ and for the energy injection 
law $E \propto t^e$ required for each model to account for the measured X-ray flux decay. 
The observer time in the above equations corresponds to the arrival-time of photons emitted by the
fluid moving at an angle $\theta = \Gamma^{-1}$ relative to the direction toward the observer, 
from where arises most of the high-energy emission received by the observer at time $t$.

 Putting together all the elements described above, we have calculated the jet final energy 
(at $t_\ended$) for the X-ray afterglows whose light-curve breaks can be explained by the a 
jet-break model with steady energy injection or an energy-injection break model (the models 
listed in the mid and lower parts of Table \ref{mod}). The jet energies for $n_\o =1$ and 
$A_*=1$, averaged over all the afterglows that can be accounted for by each model (fewer than 
indicated in Table \ref{mod}, because the jet energy calculation requires the GRB output/burst 
redshift), are listed in Table \ref{eng}. For jet-breaks models, the jet opening is determined 
by the light-curve break epoch. For energy injection-break models, the epoch of the first (last)
X-ray measurement provide an upper (lower) limit on the jet-break epoch, which yields an upper 
(lower) limit on the jet opening and, consequently, on the jet energy. Numbers in round brackets 
give the dispersion of individual jet energies (or their lower/upper limits).

\begin{table}
 \caption{ Average (and dispersion) of the logarithm of jet energy (in ergs) at the last X-ray 
   measurement, for the models in the two lower sections of Table \ref{mod}. }
\begin{tabular}{lccccccccc}
   \hline \hline
 {\sc model} &  \multicolumn{2}{c}{\sc Homog medium}  & \multicolumn{2}{c}{\sc Wind-like medium} \\
                     &  $\nu < \nu_\c$ &  $\nu_\c < \nu$ &   $\nu < \nu_\c$ &  $\nu_\c < \nu$ \\
   \hline
 \multicolumn{5}{c}{jet-break}    \\
  \hline
 J-syn-cjet & 50.4(0.5) & 49.9(0.6) & 51.1(0.6) & 49.9(0.3) \\ 
 J-syn-sjet & 50.3(0.8) & 50.1(0.7) & 51.5(0.6) & 50.2(0.4) \\
 J-iC-cjet  & 50.7(0.6) & 49.8(0.7) &  ---      & 50.0(0.3) \\
 J-iC-sjet  & 50.6(0.7) & 49.6(0.6) &  ---      & 49.7(0.3) \\
  \hline
 \multicolumn{5}{c}{energy-injection break} \\
   \hline
 EI-syn-cjet $<$ & 50.0(0.6) & 49.7(0.6) & 51.7(0.7) & 49.9(0.5) \\ 
 EI-syn-sjet $<$ & 50.8(0.7) & 50.2(0.7) & 51.7(0.6) & 50.3(0.4) \\
 EI-syn-sph  $>$ & 51.0(0.5) & 50.6(0.4) & 51.8(0.6) & 50.6(0.3) \\
 EI-iC-cjet  $<$ & 50.2(0.7) & 49.7(0.7) & 53.6(0.8) & 50.0(0.5) \\  
 EI-iC-sjet  $<$ & 50.4(0.7) & 48.9(0.8) & 55.0(1.3) & 49.5(0.6) \\ 
 EI-iC-sph   $>$ & 51.4(0.6) & 50.7(0.5) & 53.3(0.8) & 50.6(0.3) \\  
 \hline \hline
\end{tabular}
\label{eng}
\end{table}

 A potentially important result shown in Table \ref{eng} is that models where X-ray is inverse-Compton, 
the ambient medium is a wind, and X-ray is lower than the cooling frequency, can be ruled out on 
energetic grounds because, for them, the required jet total energy exceeds $10^{53.5}$ erg\footnotemark.
\footnotetext{ Magneto-hydrodynamic simulations of accretion disks (e.g. McKinney 2005, Hawley \& Krolik 2006) 
   show that the energy in relativistic matter (electromagnetic outflow) extracted from an accreting black-hole 
  increases from 0.2 (0.03) percent of the accreted mass, for a non-rotating black-hole, to 15-40 (5-20) 
  percent for a maximally rotating one. The maximal energy budget quoted here corresponds to a 20 percent 
  MHD efficiency and a $1\,M_\odot$ accretion disk.  } 
Those models lead to a high final jet energy because a substantial increase of the jet energy is required 
to account for the measured X-ray flux decay rate, which stems from that the corresponding adiabatic 
jet model yields a steep X-ray flux decay (Table \ref{alfa}: $\alpha_\x = p+1/2 = 2\beta_\x + 3/2$, for 
a conical jet), from that energy injection has a weak effect in mitigating the flux decay ($d\alpha_\x/de = 
p-2 =2\beta_\x -1$, for the spreading jet model), or a combination of those two factors (for the spherical 
outflow model).
 
 Assuming the same jet energy at the beginning of the afterglow, one can also calculate the ratios of 
jet energies among the three types of X-ray break models and outflow dynamics: 
$i)$ a narrow-jet model with an energy-injection break (models E-cjet and E-sjet in Table \ref{eng}), 
$ii)$ a jet-break model with steady energy injection (models J-cjet and J-sjet), and 
$iii)$ a wide-jet model with energy-injection break (model E-sph). 
The interesting result is that, going from type $i)$ to type $ii)$ models above increases on 
average (i.e over afterglows) the total jet energy by a factor 2-3, with another increase by a factor 
2-3 required when going from type $ii)$ to $iii)$ models. These results hold for either type of ambient medium, 
either location of the cooling frequency relative to the X-ray, and for either X-ray radiative process,
if the jet does not spread sideways, and only for inverse-Compton emission, if the jet is spreading
laterally. Thus, the most economical model for a light-curve break is that of a narrow jet with an early 
jet-break and a broken power-law energy injection origin for the X-ray light-curve break. However, these
conclusions do not hold for spreading jets if the X-ray radiation process is synchrotron. In this case,
we find that the required jet energy is about the same for all model types above. Note that these
relative energy requirements cannot be deduced from the average jet energies listed in Table \ref{eng}.
Here, we are referring to averages over afterglows of jet energy ratios between two models that explain
the same afterglow break, while Table \ref{eng} gives jet energies for a given model averaged over all 
the afterglows that can be explained by that model.

\subsection{Loose ends}

 The first caveat/weakness of the model testing presented above was already discussed: 
the energy-injection break origin for X-ray light-curve breaks is only weakly falsifiable 
by a comparison between the model $\alpha-\beta$ closure relation and the observed flux 
power-law decay index $\alpha$ and spectral slope $\beta$.
The only way for this model to fail is an X-ray flux that decays steeper than expected 
for an adiabatic shock, as energy injection can only mitigate the afterglow flux decay. 

 Furthermore, a comparison of the model light-curve breaks against observations is needed 
to validate possible X-ray light-curve break models, as the sharpness of X-ray light-curve 
breaks varies among afterglows. The majority of the well-sampled light-curve breaks used
here are sharp, lasting less than a factor two in time. Quite likely, the sharpest breaks 
cannot be accounted by a wind-like medium, whatever is the break mechanism, owing to the 
slower jet deceleration produced by a decreasing external density (Kumar \& Panaitescu 2000). 
 Without taking into account such limitations, testing only the ability of some models
to account for the observed flux power-law decay indices will overestimate the fraction of 
afterglows that can be explained by a given model.

 We emphasize the importance of using afterglows whose light-curves have been sampled over a 
sufficiently long time before and after the break, so that the asymptotic flux decay indices 
can be measured accurately. Otherwise, shortly-monitored light-curves with breaks can lead to an 
underestimation of the break magnitude $\delta \alpha = \alpha_{\rm x2}-\alpha_{\rm x1}$, which 
will favour a steady energy injection interpretation, because its effect is to reduce the break 
magnitude $\delta \alpha$. We note that eighty percent of the power-law segments used in this 
work last a decade or more in time.

\subsubsection{Optical afterglows}

 Inclusion of the optical light-curves could provide a further test to the origin of X-ray 
light-curve breaks (e.g. Liang et al 2007; Curran et al 2009; Liang et al 2008) 
in the two models discussed above (jet-break and energy-injection break). 
Both models yield {\sl achromatic} light-curve breaks, appearing at the same time in all
light-curves. For those afterglows where it can be established that the optical and X-ray
emissions arise from the same source, either because the light-curve breaks are achromatic
or because the optical and X-ray spectra are on the same power-law continuum, testing the
light-curve break origin and determining the model details (location of cooling frequency,
ambient medium stratification, radiation process, outflow dynamics) can be done by comparing
the optical flux decay indices $\alpha_\o$ with the model expectations, for the energy injection
parameter $e$ required by the X-ray flux decay indices $\alpha_\x$. 

 In GRBlog database of optical light-curves (Quimby, McMahon \& Jeremy 2004), we find five 
afterglows (050401, 061121, 070420, 090424, 090618) listed in Table \ref{ag1} with chromatic 
X-ray breaks, hence their X-ray and optical emissions arise from different parts of the outflow, 
and eight with achromatic breaks, thus a comprehensive test of the X-ray break models cannot be 
performed at this time. We note that, of those 8 afterglows with achromatic breaks, two (060729, 
081029) have $\alpha_\x \simeq \alpha_\o$, for five (050922C, 051109, 060614, 080710, 090424) 
$0.25 < \alpha_\x - \alpha_\o  < 0.5$, and for one (090510) $\alpha_\x - \alpha_\o  \simeq 1$. 
 
 In general, $\alpha_\x = \alpha_\o$ (as for 060729) is expected if the cooling frequency is not 
between the optical and X-ray and $\alpha_\x - \alpha_\o = -(1/2) d\log \nu_\c /d \log t$ if 
$\nu_\c$ is in between. According to Table \ref{spek}, a decreasing $\nu_\c$
(which leads to $\alpha_\x > \alpha_\o$), is obtained only for a homogeneous medium, thus all
six afterglows with $\alpha_\x - \alpha_\o \geq 0.25$ require such an ambient medium. 
Energy injection accelerates the evolution of $\nu_\c$ and increases the index 
difference $\alpha_\x - \alpha_\o$ above the expectation for an adiabatic outflow:
$\alpha_\x - \alpha_\o  = (e+1)/4$ for a spherical/conical outflow and $\alpha_\x - \alpha_\o = e/3$ 
for a spreading jet. Thus, the energy injection indices $e$ in the range $(0.3,1)$ required by the
X-ray decays of those six afterglows imply that $0.3 < \alpha_\x - \alpha_\o < 0,5$, consistent 
with what is measured for five of them, while the large $\alpha_\x - \alpha_\o$ measured for 090510 
indicates that the optical and X-ray are from different radiative processes (synchrotron and 
inverse-Compton, respectively).

\subsubsection{Reverse shock}

 In this work, we have considered only the forward-shock emission and ignored the emission
from the reverse-shock, which energizes the GRB/post-GRB ejecta, the agent of energy injection 
into the blast-wave. The two shocks dissipate equally the kinetic energy of the ejecta, 
which indicates that the reverse-shock could make a comparable contribution to the afterglow emission. 

 For a short-lived ejecta source, it can be shown from the kinematics of the catch-up between 
the GRB ejecta and the decelerating forward-shock that the ratio of the Lorentz factor 
$\Gamma_\i$ of the ejecta arriving at the forward-shock at time $t$ to that of the swept-up 
ambient medium $\Gamma(t)$ is $\Gamma_\i/\Gamma(t) =[(4-s)/(1+e)]^{1/2} \gta 1$ (evidently,
the ratio of those two Lorentz factors must be above unity, but cannot be much above unity,
else the catch-up would have occurred earlier). In this case, 
the reverse shock is only mildly relativistic and the ejecta electrons accelerated by the 
reverse-shock could be much less energetic than the electrons behind the forward-shock, 
making the emission of the latter dominant at X-ray photon energies. 

 For a long-lived central source, the contrast between the Lorentz factor of the unshocked
outflow and that of the forward-shock can be arbitrarily large, the reverse-shock could be
highly relativistic, provided that the ejecta are not too dense (e.g. Sari \& Piran 1995). 
The decay of the reverse-shock 
flux can be calculated in a way similar to that shown here for the forward-shock; however, 
the parametrization of the reverse-shock emission is not that simple because 
$i$) its dynamics depends on the Lorentz factor of the ejecta and their density, and 
$ii$) its peak flux depends on the total ejecta mass having entered the reverse-shock. 
Thus, the calculation of the reverse-shock emission requires at least two parameters that
quantify the energy and mass mass injection into the blast-wave, leading to a model even 
less constrainable (observationally) than the forward-shock (which requires only one parameter,
$e$, for the energy injection law).

\section{Conclusions}

 By comparing the flux decay index ($\alpha$) with the spectral slope ($\beta$) measured for a set 
of 98 well-sampled Swift X-ray afterglows with the expectations for the forward-shock model, we find that
about a third of those afterglows display a break that could be the traditional `jet-break' claimed to 
have been observed in a dozen pre-Swift optical afterglows (\S\ref{adb}). 
For this test of {\sl adiabatic} jet-breaks, we have considered various details of the forward-shock 
model: homogeneous or wind-like ambient media, spectral cooling frequency higher or lower than the X-rays, 
synchrotron or inverse-Compton as the dominant X-ray process, a collimated conical or spreading outflow. 

 If a steady energy injection into the forward-shock occurs, then the jet-break model may account for 
88 percent of the 117 light-curve breaks or, else put, for at least one break in 87 percent of the 98 
afterglows selected here (\S\ref{jbei}). This vast improvement over the adiabatic jet model suggests that 
energy injection could be a prevalent process in GRB afterglows.

 The remaining 12 percent of breaks that cannot be explained by a jet-break undergoing a steady energy 
injection, and the 19 afterglows with two breaks (only one being, at most, a jet-break), indicate that 
some X-ray breaks may originate from a change in the rate at which energy is added to the forward-shock 
(an energy-injection break), and it is possible that some (many ?) of the identified jet-breaks are also 
energy-injection breaks (\S\ref{eibr}).
It should be recognized that the complete success of the energy-injection break model over the traditional 
jet-break model could be due mostly to that the latter model is over-constrained (observations provide 
two constraints: flux pre- and post-break power-law decay indices -- for one model parameter: the energy 
injection power-law exponent), while the former model is not.  

 We find a tentative correlation between the GRB output and the afterglow X-ray flux decay rate, 
more energetic bursts being followed by faster decaying afterglows (\S\ref{fag}). 
If energy injection is a common process in GRB afterglows, then the slower decaying afterglows 
should be identified with a stronger total energy injection. This means that the GRB energy --
afterglow (X-ray flux) decay rate correlation should lead to an anticorrelation of the GRB 
output with the increase of the outflow kinetic energy during the afterglow phase, i.e. more 
energetic bursts should be followed by afterglows with less energy being injected. 
This anticorrelation is, indeed, observed for both the jet-break and the energy injection 
break models for X-ray light-curve breaks, and for various model details (dynamics, X-ray 
afterglow emission process, radial stratification of the ambient medium). If the GRB progenitor 
is a long-lived source of relativistic ejecta, the above anticorrelation implies an anticorrelation 
of the kinetic energy of the leading outflow (producing the burst) with that of the following
ejecta (which inject energy in the blast-wave during the afterglow phase), but the total
ejecta energy (burst plus afterglow) is far from being universal.

 The energy budget required by most of the possible variants of the forward-shock model ranges
from $10^{49}$ to $10^{52}$ ergs, and only variants involving inverse-Compton emission in the
X-rays and a wind-like medium can be excluded on energetic grounds because, in those models, 
the required jet-energy exceeds $10^{53}$ erg (\S\ref{jeten}). Considering both models for light-curve
breaks (jet-break with steady energy-injection and energy-injection break), we find that the
most energetically economical model is that of a narrow jet whose jet-break occurs early on
(before first X-ray measurement) with the light-curve break being an energy-injection break, 
followed by a model where the X-ray light-curve break is a jet-break, with the most "wasteful" model 
being that of a wide jet whose break occurs late (after last X-ray measurement), the X-ray break being
an energy-injection break. However, the average (over afterglows) energy ratio among pairs of these 
three break models is only 2--3. 


\section*{Acknowledgments}
 This work was supported by an award from the Laboratory Directed Research and Development program 
 at the Los Alamos National Laboratory and made use of data supplied by the UK Science Data Center 
 at the University of Leicester, by the ``GRB log" site for optical light-curves (http://grblog.org/grblog.php), 
 and by the GRB afterglow repository of GCN circulars (http://www.mpe.mpg.de/$\sim$jcg/grbgen.html).

\end{document}